\documentclass[12pt]{article}
\usepackage{a4wide}
\usepackage{graphicx}
\usepackage{tabularx}
\newcommand{\be}{\begin{equation}}
\newcommand{\ee}{\end{equation}}
\newcommand{\ba}{\begin{eqnarray}}
\newcommand{\ea}{\end{eqnarray}}
\usepackage{color}

\makeatletter
\def\hlinewd#1{%
\noalign{\ifnum0=`}\fi\hrule \@height #1 %
\futurelet\reserved@a\@xhline}
\makeatother
\begin{document}
\begin{titlepage}
\begin{flushright}
\end{flushright}
\begin{center}
{\Large {\bf Kaon semi-leptonic form factor at zero momentum transfer in finite volume }} \\
\vspace{1cm}
{\bf Karim Ghorbani$^{1}$ and Hossein Ghorbani$^{2}$} \\[1cm]
\it{{$^{1}$Physics Department, Faculty of Sciences, Arak University, Arak 38156-8-8349, Iran \\
$^{2}$School of Particles and Accelerators, Institute for Research in Fundamental Sciences (IPM),
P.O. Box 19395-5531, Tehran, Iran}}
\end{center}

\begin{abstract}
Using Chiral Perturbation Theory, we obtain the kaon
semi-leptonic vector form factor in finite volume
at a generic momentum transfer, $q^2$, up to one loop order.
At first we confirm the lattice observation that the contribution
of the heavy Pseudo-Goldstone boson in the finite volume corrections
at zero momentum transfer is unimportant.
We then evaluate the form factor at $q^2=0$ numerically and
compare our results with the present lattice data.
It turns out that our ChPT results are comparable
with the lattice data to some extend.
The formula for the finite volume corrections
obtained for the form factor at momentum transfer $q^2$,
provides a tool for lattice data in order to
extrapolate at large lattice size.

\end{abstract}
\vfill
\vspace{3.5cm}

\end{titlepage}

\section{Introduction}
\label{intro}
Lattice QCD calculations have improved significantly in recent years
such that extraction of the parameters of the standard model are feasible,
see for example \cite{Boyle:2013,Bazavov:2012cd,Kaneko:2012cta,Kronfeld:2012uk,Na:2011mc,Boyle:2010bh,Boyle:2007qe,Lubicz2009,Dawson2006}.
For a review on the lattice result on the low energy particle physics
one may consult reference \cite{Colangelo:2010et}.
Among these parameters, the precision evaluation of the CKM matrix
element, namely $|V_{us}|$ is important from the vantage point
of finding new physics footprint in the unitarity requirement of the first
row of the CKM matrix \cite{Cabibbo:1963,Kobayashi:1973}.
The main uncertainty in this unitarity relation is due to $|V_{us}|$.
From the measurements of the decay rate of the semi-leptonic
kaon decay ($K \to \pi e \nu$), the so-called $K_{l3}$ decay,
one can only determine the combination
$|V_{us}|f_{+}(0)$, where the quantity $f_{+}(0)$ is
the relevant vector form factor at zero momentum transfer.
The average of measurements from different modes of $K_{l3}$
is provided by PDG(2012)\cite{Beringer:1900zz}
\begin{equation}
\label{PDG}
|V_{us}|f_{+}(0) = 0.21664 \pm 0.00048\,.
\end{equation}
At our disposal are two approaches to determine $|V_{us}|$.
On one hand, lattice QCD  (LQCD) provides the vector form factor
by numerical evaluation of the relevant functional integral
of QCD, for instance in \cite{Boyle:2007qe} it is found
\begin{equation}
\label{PDG}
f_{+}(0)_{LQCD} = 0.9644 \pm 0.0033  \pm 0.0034 \pm 0.0014,
\end{equation}
where the first error is statistical and the second and
third errors are the size of the systematic errors.
As it is noted in \cite{Boyle:2007qe}, the uncertainty
in the lattice data is dominated
by the statistical, chiral\footnote{Since the collaborations RBC+UKQCD
and MILC are now simulating at the physical point, the chiral
extrapolation will be unnecessary soon.}
and $q^2$ extrapolation\footnote{When partially twisted boundary condition is used the $q^2$ extrapolation
is entirely absent.}.

On the other hand, there is the application of an effective
field theory, namely, chiral perturbation theory (ChPT).
In this framework, the evaluation of $f_{+}(t)$
at one loop order is done by Gasser and Leutwyler \cite{Gasser-Leutwyler}.
According to the Ademollo-Gatto theorem, at one loop order, $f_{+}(0)$ is free
from order $p^4$ low energy constants $L^{r}_{i}$, and therefore
at this order, chiral effective theory predicts the form factor at zero momentum transfer
unambiguously in terms of logarithmic corrections.
There exist higher order ChPT based works
\cite{Bijnens-Talavera,Jamin2004,Cirigliano2005,Bijnens-ghorbani2007}
which indicate the dominance of the two loop quantum corrections to $f_{+}(0)$.
It should be pointed out, however, that ChPT result for $f_{+}(0)$ at two loop order
contains a correction as $\Delta f = -8~(M^2_K-M^2_\pi)^2~(C^{r}_{12}+C^{r}_{34})$
arising from higher order local operators.
The combination of low energy constants $(C^{r}_{12}+C^{r}_{34})$, is not predicted by
ChPT.

However, an estimate of this correction based on a quark model is already discussed by
Leutwyler and Roos in \cite{Leutwyler:1984} which is not satisfactory for the high precision
extraction of $V_{us}$ from experiment.
Today, instead of quark model, the combination of $p^6$ constants can be obtained from
Lattice QCD, see \cite{Bazavov2012} for detailed discussion. However, to determine
the $p^6$ constants within Lattice QCD, the $p^4$ constants are needed as input.

In this work we have made an attempt in order to make an estimate for the finite volume
effects of $f_{+}(0)$ with non-vanishing spatial momentum transfer
in the framework of ChPT. These results can be considered as
a guideline for lattice practitioners when they do the large volume extrapolation.
In the present article we calculate the form factors at finite volume for
a generic momentum transfer which is a generalization of a previous work in \cite{Kgh2011}.
It was shown in \cite{Kgh2011} that the scalar form factor
at the maximum momentum transfer acquires unexpected large
finite volume correction
for typical pion masses used in lattice calculations.

The main motivation behind
this work is actually to figure out the usefulness of ChPT application in
finite volume for $f_{+}(0)$. Many works in the literature can be found
on employing ChPT to estimate the systematic errors.
The finite size effects on the pion mass and pion decay constant
in \cite{Colangelo2006,Colangelo2005,Colangelo2004} and on
quark vacuum expectation values in \cite{Kgh2006}
are examples for cases in which external momenta are not involved.
There is also a work for the meson matrix element in finite
volume, with non-zero external spatial momentum in \cite{Bunton}.

The rest of the article is organized as follows.
In Sec.~\ref{chpt} chiral perturbation theory
is briefly introduced. The next section introduces
the strangeness-changing semi-leptonic kaon decay.
The hadronic matrix element for the process
in a tensor form at one loop order is recapitulated in
Sec.~\ref{analytical} and pion mass dependence of $f_{+}(0)$ is
studied in infinite volume.
A short introduction to the application of
ChPT in finite volume is given in Sec.~\ref{chpt-application}
and all needed Feynman integrals in finite volume
are calculated anew in the Appendix A.
In Sec.~\ref{deltaf} we present our analytical formula for the finite box correction
at momentum transfer, t.
Finally, in Sec.~\ref{formfactor-fv}
our numerical results and comparison with lattice data are presented.
We finish up with a conclusion.

\section{SU(3) chiral perturbation theory}
\label{chpt}
At low energies Quantum Chromo Dynamics (QCD) becomes a strongly coupled theory
therefore, the standard perturbation approach is no longer applicable.
Chiral perturbation theory (ChPT) is an effective field theory to
study the strong interactions at low energy. Spontaneously chiral symmetry breaking
in QCD gives rise to Pseudo-Goldstone mesons which are considered as
dynamical degrees of freedoms in the effective theory. ChPT is emerged in its
modern formalism in a paper by Weinberg \cite{Weinberg0} and developed latter on by
Gasser and Leutwyler to higher orders \cite{GL0,GL1}.
External momentum, $p^2$ and quark masses, $m_{q}$ are the generic
expansion parameters.
At the lowest order SU(3) chiral lagrangian
contains two terms and takes on the form \cite{Weinberg0}
\be
\label{Lp2}
{\cal L}_{2} = \frac{F_{0}^{2}}{4} \langle u_{\mu} u^{\mu}+ \chi_{+} \rangle,
\ee
where $F_{0}$ is the pion decay constant at chiral limit
and $\langle...\rangle$ = $ \mathrm{Tr}_F\left(...\right)$
stands for the trace over the flavors.
We introduce the matrices $u^{\mu}$ and $\chi_{\pm}$ as the following
\ba
u_{\mu} = i u^{\dag}D_{\mu}U u^{\dag} = u_{\mu}^{\dag}  \,, \quad u^{2} = U,
\nonumber\\
\chi_{\pm} = u^{\dag} \chi u^{\dag} \pm u\chi^{\dag}u.
\ea
We could parameterize $\chi$ in terms of scalar and pseudo-scalar
external densities but in this work it is enough to set
\ba
\chi
 = 2B_{0}\, \left( \begin{array}{ccc}
\displaystyle m_{u} &   \\
    &\displaystyle m_{d} &  \\
 &   &\displaystyle m_{s}
\end{array}  \right).
\ea
The matrix $U \in SU(3)$ incorporates the octet of the light
pseudo-scalar mesons
\be
U(\phi) = \exp(i \sqrt{2} \phi/F_0)\,,
\ee
where
\ba
\phi (x)
 = \, \left( \begin{array}{ccc}
\displaystyle\frac{ \pi_3}{ \sqrt 2} \, + \, \frac{ \eta_8}{ \sqrt 6}
 & \pi^+ & K^+ \\
\pi^- &\displaystyle - \frac{\pi_3}{\sqrt 2} \, + \, \frac{ \eta_8}
{\sqrt 6}    & K^0 \\
K^- & \bar K^0 &\displaystyle - \frac{ 2 \, \eta_8}{\sqrt 6}
\end{array}  \right) .
\ea
We can obtain the relation $m_{\pi}^2 = B_{0}(m_{u}+m_{d})$ from the
lowest order lagrangian. This relation allows us to take quark masses
of order $p^2$.
In the covariant derivatives, external fields are defined as
\ba
D_{\mu} U = \partial_{\mu} U - i r_{\mu}U +iUl_{\mu}.
\ea
The left- and right-handed external fields are expressed
by $l_{\mu}$ and $r_{\mu}$ respectively.
For the process we consider in this paper we just need to set
\begin{eqnarray}
l_{\mu} = \frac{g_{2}}{\sqrt{2}} \left( \begin{array}{ccc}
\displaystyle  & V_{ud} W_{\mu}^{+} & V_{us} W_{\mu}^{+} \\
V_{ud}^{\ast} W_{\mu}^{-}   &  &  \\
V_{us}^{\ast} W_{\mu}^{-}   &  &
\end{array}\right),
r_{\mu} = 0.
\nonumber\\
\end{eqnarray}
The weak coupling constant, $g_{2}$, is given in terms of Fermi constant and
$W$ mass by the relation $g_{2}^2 = 4 \sqrt{2} G_{F} m_{W}^2$.
The next to leading order effective lagrangian provided by \cite{GL0,GL1}
contains twelve independent operators
\begin{eqnarray}
\label{lagL4}
{\cal L}_4&&\hspace{-0.5cm} =
L_1 \langle u_{\mu} u^{\mu} \rangle^2
+L_{2} \langle u_{\mu} u^{\nu} \rangle \langle u^{\mu} u_{\nu} \rangle
+L_{3} \langle u_{\mu} u^{\mu} u_{\nu} u^{\nu} \rangle
+L_{4} \langle u_{\mu} u^{\mu} \rangle \langle \chi_{+} \rangle
\nonumber\\&&\hspace{-0.1cm}
+L_{5} \langle u_{\mu} u^{\mu} \chi_{+} \rangle
+L_{6} \langle \chi_{+} \rangle^2+ L_{7} \langle \chi_{-} \rangle^2
+\frac{1}{4}(2L_{8}+L_{12})\langle \chi_{+}^2 \rangle
\nonumber\\&&\hspace{-0.1cm}
+\frac{1}{4}(2L_{8}-L_{12})\langle \chi_{-}^2 \rangle
-iL_{9} \langle f^{\mu \nu}_{+} u_{\mu} u_{\nu} \rangle
+\frac{1}{4}(L_{10}+2L_{11}) \langle f_{+\mu \nu} f_{+}^{\mu \nu}  \rangle
\nonumber\\&&\hspace{-0.1cm}
-\frac{1}{4}(L_{10}-2L_{11}) \langle f_{-\mu \nu} f_{-}^{\mu \nu}  \rangle,
\end{eqnarray}
where $L_{i}$ are the low energy constants which are obtainable phenomenologically
and the field strength tensor is defined as
\begin{eqnarray}
f_{\pm}^{\mu \nu}&&\hspace{-0.5cm} = u F^{\mu \nu}_{L} u^{\dag} \pm u^{\dag} F_{R}^{\mu \nu} u,
\nonumber\\
F^{\mu \nu}_{L}&&\hspace{-0.5cm} = \partial^{\mu}l^{\nu}-\partial^{\nu}l^{\mu}-i[l^{\mu},l^{\nu}],
\nonumber\\
F^{\mu \nu}_{R}&&\hspace{-0.5cm} = \partial^{\mu}r^{\nu}-\partial^{\nu}r^{\mu}-i[r^{\mu},r^{\nu}].
\end{eqnarray}

\section{The definition of the $K \to \pi$ form factors}
\label{kl3defin}
Semileptonic weak decays of charge and neutral kaon known as $K_{l3}$ are:
\begin{equation}
K^{+}(p) \to \pi^{0}(p^{\prime}) l^{+}(p_{l}) \nu_{l}(p_{\nu}) \,,
\end{equation}
\begin{equation}
K^{0}(p) \to \pi^{-}(p^{\prime}) l^{+}(p_{l}) \nu_{l}(p_{\nu}),
\end{equation}
where subscript $l$ indicates electron or muon.
There are two other processes which are the charge conjugate modes of the decays above.
The matrix element for these processes, e.g. the neutral mode,
\begin{equation}
{\cal K} = \frac{G_{F}}{\sqrt{2}} V^{\ast}_{us} J^{\mu} {\cal M}_{\mu} (p^{\prime},p),
\end{equation}
consists of two parts, keeping only the vector contributions.
One part which defines the purely leptonic current
\begin{eqnarray}
J^{\mu} &=& {\bar u}(p_{\nu}) \gamma^{\mu}(1-\gamma_{5}) v(p_{l}),
\end{eqnarray}
and the other part being our concern in this paper, incorporates the hadronic kaon-pion weak transition
\begin{eqnarray}
{\cal M}_{\mu}(p^{\prime},p) &=& \hspace{0.1cm}<\pi^{-}(p^{\prime})| {\bar s} \gamma_{\mu} u(0)|K^{0}(p)>.
\end{eqnarray}
The hadronic matrix element is generally defined by
\begin{eqnarray}
<\pi^{-}(p^{\prime})| {\bar s} \gamma_{\mu} u(0)|K^{0}(p)> &=&
\frac{1}{\sqrt{2}}[(p+p^{\prime})_{\mu} f^{K^{0}\pi^{-}}_{+}(t)
+(p-p^{\prime})_{\mu}f^{K^{0}\pi^{-}}_{-}(t)].
\nonumber\\
\end{eqnarray}
A similar definition can be provided for the charge kaon.
The two $K_{l3}$ vector form factors $f^{K^{0}\pi^{-}}_{\pm}(t)$ depend on
the four-momentum squared, t, transferred to the leptons:
\begin{equation}
t = (p-p^{\prime})^{2} = (p_{l}+p_{\nu})^2.
\end{equation}
The so-called scalar form factor as the S-wave projection of the matrix element can be defined as
\begin{equation}
\label{scalarform}
f_{0}(t) = f_{+}(t)+ \frac{t}{m_{K}^2-m_{\pi}^2} f_{-}(t).
\end{equation}
Given the definitions for the vector and scalar form factors, it is possible to
obtain these dynamical low energy quantities in terms of temporal and
spatial parts of the hadronic matrix element, namely, ${\cal M}_{\mu}(p^{\prime},p)$.
For the vector form factor we find
\begin{equation}
\label{vectorform}
 f_{+}(t) = \frac{(p_{i}-p_{i}^{\prime}){\cal M}_{0}-(E_{p}-E_{p^{\prime}}){\cal M}_{i}}
          {\sqrt 2 (E_{p^{\prime}}p_{i}-E_{p}p_{i}^{\prime})},
\end{equation}
and for the scalar form factor we obtain
\begin{equation}
\label{scalarform}
 f_{0}(t) = f_{+}(t) [1+\frac{t}{M_{K}^2-M_{\pi}^2} \frac{(E_{\vec p^{\prime}}p_{i}-\vec E_{\vec p}p_{i}^{\prime})}{(E_{\vec p}p_{i}^{\prime}- E_{\vec p^{\prime}}p_{i})}    \frac{(p_{i}+p_{i}^{\prime}){\cal M}_{0}-(E_{\vec p}+E_{\vec p^{\prime}}){\cal M}_{i}}
                                {(p_{i}-p_{i}^{\prime}){\cal M}_{0}-(E_{\vec p}-E_{\vec p^{\prime}}){\cal M}_{i}} ],
\end{equation}
where ${\cal M}_{0}$ and ${\cal M}_{i}$ are respectively, the temporal and the spatial components of the weak vector
current. The Kaon and Pion energies are given by $E_{\vec p}= \sqrt{M_{K}^2+{\vec p}^2}$ and
$E_{\vec p^{\prime}}= \sqrt{M_{\pi}^2+\vec p^{\prime^2}}$, respectively.
The relations above are useful when we look at the form factors in finite space
in subsequent sections.

\section{The weak matrix element and $f^{\infty}_{+}(0)$ }
In order to evaluate the form factors in finite volume one needs
the matrix element in a tensor form.
To this end in \cite{Kgh2011} the hadronic matrix element at one
loop order in the isospin limit is found
\begin{eqnarray}
\label{matrix}
{\cal M}(p^{\prime},p).\epsilon &=&
\frac{1}{F_{\pi}^2} \Big [ 2 q^{2} L_{9} + [\frac{3}{8}A(m_{\pi}^2)+\frac{3}{8}A(m_{\eta}^2)
+\frac{3}{4}A(m_{K}^2)]r.\epsilon
-[\frac{3}{2}B_{\mu \nu}(m_{\pi}^2,m_{K}^2,q^2)
\nonumber\\&&\hspace{-0.1cm}
+\frac{3}{2} B_{\mu \nu} (m_{K}^2,m_{\eta}^2,q^2)]r^{\nu} \epsilon^{\mu}
+[ -2(m_{K}^2 - m_{\pi}^2) L_{9}
+4(m_{K}^2 - m_{\pi}^2) L_{5} + \frac{1}{2} A(m_{\eta}^2)
\nonumber\\&&\hspace{-0.1cm}
-\frac{5}{12}A(m_{\pi}^2)
+\frac{7}{12} A(m_{K}^2)] q.\epsilon
+B(m_{\pi}^2,m_{K}^2,q^2)(\frac{5}{12}q^2
-\frac{5}{12} m_{K}^2
-\frac{1}{12}m_{\pi}^2) q.\epsilon
\nonumber\\&&\hspace{-0.1cm}
+B(m_{K}^2,m_{\eta}^2,q^2)(\frac{1}{4}q^2
-\frac{7}{12} m_{K}^2
+\frac{1}{12}m_{\pi}^2) q.\epsilon
-[\frac{5}{6}B_{\mu\nu}(m_{\pi}^2,m_{K}^2,q^2)
\nonumber\\&&\hspace{-0.1cm}
+\frac{1}{2}B_{\mu\nu}(m_{K}^2,m_{\eta}^2,q^2)]q^{\mu} \epsilon^{\nu}
+B_{\mu}(m_{\pi}^2,m_{K}^2,q^2)
[\frac{3}{4}(p+p^{\prime})^{\mu} q.\epsilon+\frac{5}{12}q^{\mu} q.\epsilon
\nonumber\\&&\hspace{-0.1cm}
+\frac{5}{6}m_{K}^2 \epsilon^{\mu}
+\frac{1}{6} m_{\pi}^2 \epsilon^{\mu}
-\frac{5}{6} q^{2} \epsilon^{\mu}]
+B_{\mu}(m_{K}^2,m_{\eta}^2,q^2)
[\frac{3}{4}(p+p^{\prime})^{\mu} q.\epsilon
\nonumber\\&&\hspace{-0.1cm}
+\frac{1}{4}q^{\mu} q.\epsilon
+\frac{7}{6}m_{K}^2 \epsilon^{\mu}-\frac{1}{6} m_{\pi}^2 \epsilon^{\mu}
-\frac{1}{2} q^{2}  \epsilon^{\mu}] \Big ],
\end{eqnarray}
where
\begin{eqnarray}
r = p^{\prime}+p \,,  &&\hspace{.5cm}  q = p - p^{\prime} \,,
\end{eqnarray}
and $\epsilon$ is the polarization four-vector of the $W$ boson.
In the expression above scalar integrals $A$ and $B$, as well as tensor integrals
$B^{\mu}$ and $B^{\mu \nu}$ are introduced in the Appendix B.
In infinite volume, we can reduce the tensor integrals to scalar integrals
by relations given in the Appendix B.
It is then possible to obtain the vector form factors following a renormalization
program and in the end, the known vector form factors can be reproduced.
At zero momentum transfer, the vector form factor at infinite volume reads \cite{Leutwyler:1984},
\begin{eqnarray}
\label{fzero}
f_+^{\infty}(0) &&\hspace{-0.1cm}  =
\frac{1}{F_\pi^2}\Big[\frac{3}{8}\,\overline{A}(m_{\pi}^2)+\frac{3}{4}\,\overline{A}(m_{K}^2)
+\frac{3}{8}\,\overline{A}(m_{\eta}^2)
-\frac{3}{2}\,\overline{B}_{22}(m_{\pi}^2,m_{K}^2,0)
\nonumber\\&&\hspace{-0.1cm}
-\frac{3}{2}\,\overline{B}_{22}(m_{K}^2,m_{\eta}^2,0)\Big]\,,
\end{eqnarray}
where
\begin{equation}
 \overline{A}(m^2) = -\frac{m^2}{16\pi^2} log(m^2/\mu^2)
\end{equation}
\begin{equation}
 \overline{B}_{22}(m^2,M^2,0)=\frac{1}{4}\frac{m^2\overline{A}(m^2)-M^2\overline{A}(M^2)}{m^2-M^2}
+\frac{m^2+M^2}{128\pi^2}.
\end{equation}
\label{analytical}

\begin{figure}
\begin{center}
 \resizebox{0.7\textwidth}{!}{%
 \includegraphics[angle=-90]{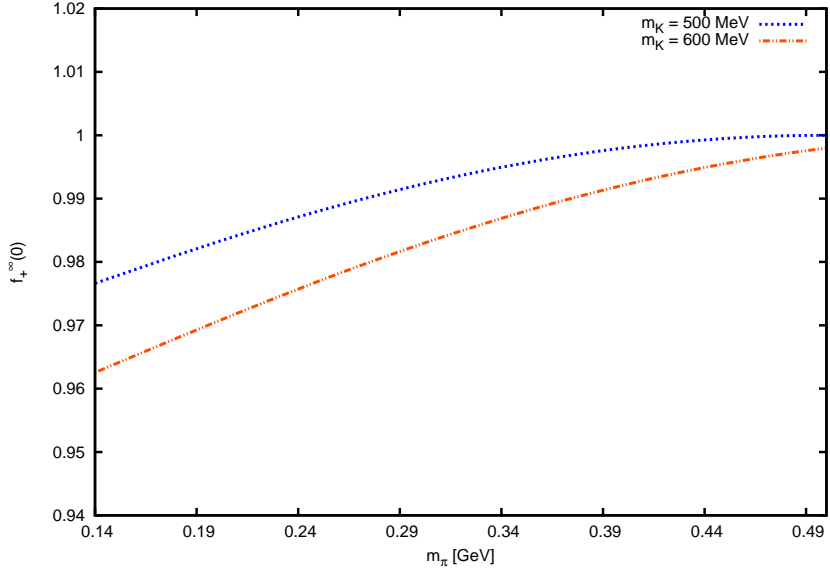}}
\caption{The form factor, $f^{\infty}_{+}(0)$, as a function
of the pion mass is shown for two different values of the kaon mass.}
\label{infinite-mass}
\end{center}
\end{figure}
It is worth mentioning that for $f^{\infty}_{+}(0)$ at next-to-leading order,
the dependency on the renormalization scale cancels.
We also compute the vector form factor, $f^{\infty}_{+}(0)$, numerically
in terms of the pion mass with choosing the physical value for the pion
decay constant, $F_{\pi} = 0.092$ GeV.
For the eta mass we use the
Gell-Mann-Oakes-Renner (GMOR) relation, $m_{\eta}^2 = (4 m_{K}^2-m_{\pi}^2)/3$.
The results shown in Fig.~\ref{infinite-mass} indicate
sizable variation of the form factor at small pion mass for
two different values of the kaon mass.
Moreover, the temporal and spacial parts of the hadronic matrix element, ${\cal M}_{\mu}(p^{\prime},p)$,
are available from Eq.~\ref{matrix}.
These are needed in our evaluation
of the form factors in finite volume as defined in Eq.~\ref{vectorform}
and Eq.~\ref{scalarform}. The relevant Feynman integrals
in finite volume are calculated in the Appendix A.

\section{ChPT application in finite lattice box}
\label{chpt-application}
In lattice QCD, simulations can be performed in a cubic volume (V=$L^3$)
with periodic boundary conditions imposed on the hadronic fields
\begin{eqnarray}
\phi(\vec x) = \phi (\vec x + {\vec n} L)  \,.
\end{eqnarray}
Therefore the three-vector momenta of hadrons become discrete
\begin{eqnarray}
\vec p = \frac{2 \pi}{L} \vec n \,,
\end{eqnarray}
where, $\vec n$  is a three dimensional vector with integer components
$(n_{x},n_{y},n_{z})$.
The application of chiral perturbation theory to study finite
volume effects are introduced in original works by Gasser and
Leutwyler, see \cite{Gasser1986finite1,Gasser1987finite2,Gasser1987finite3}
for detailed discussion in this regard. It is important to note that
with the periodic boundary condition, the effective lagrangian
in finite volume is the same as the one in infinite volume.
Finite volume corrections get their effects from modification
of the hadron propagation in space-time.
Given the quantization of the momenta
in finite volume, the two-point correlation function becomes
\begin{equation}
G_{V} = \frac{1}{L^3} \sum_{\vec{p}} \int \frac{dp^0}{2\pi} G(p^0,\vec{p})\,,
\end{equation}
where, $G(p^0,\vec{p})$ is the two-point Green function in infinite volume.
Our power counting quantity in finite volume calculations is the quantity $m_{\pi}L$
where it turns out that the zero mode of the pion filed is not strongly
coupled if the condition $m_{\pi} L >>1$ is fulfilled.
This is the so-called {\it p-regime}.
In addition, ChPT gives reliable results when $F_{\pi} L >>1$.

\section{Analytical result for $\Delta f_{+}(t)$}
\label{deltaf}
In section \ref{kl3defin} the kaon vector form factor is found in terms
of temporal and special parts of the hadronic matrix element.
Correspondingly, finite volume correction of the form factor
can be readily found as
\begin{eqnarray}
\Delta f_{+}(t) = \frac{(p_{K}-p_{\pi}){\Delta \cal M}_{0}-(E_{K}-E_{\pi}){ \Delta \cal M}_{i}}
          {\sqrt 2 (E_{\pi}p_{K}-E_{K}p_{\pi})}\,,
\end{eqnarray}
where, $p_{K}$ and $p_{\pi}$ are respectively, kaon and pion momenta along the $x$ direction.
With the hadronic matrix element available in its tensor form in section \ref{analytical} we finally obtain
\begin{eqnarray}
\Delta f_{+}(t) = \frac{p_{K}-p_{\pi}}{\sqrt 2 (E_{\pi}p_{K}-E_{K}p_{\pi})F_{\pi}^2}
(  [-\frac{1}{24}\Delta A(m_{\pi},L)+\frac{7}{8} \Delta A(m_{\eta},L)+\frac{4}{3}\Delta A(m_{K},L)]E_{K}
\nonumber\\&&\hspace{-14cm}
+[\frac{29}{24}\Delta A(m_{\pi},L)-\frac{1}{8}\Delta A(m_{\eta},L)+\frac{7}{12}\Delta A(m_{K},L)]E_{\pi}
-[\frac{7}{3}\Delta B^{00}(m_{\pi}^2,m_{K}^2,t)
\nonumber\\&&\hspace{-14cm}
+2\Delta B^{00}(m_{K}^2,m_{\eta}^2,t)]E_{K}
-[\frac{2}{3}\Delta B^{00}(m_{\pi}^2,m_{K}^2,t)+\Delta B^{00}(m_{K}^2,m_{\eta}^2,t)]E_{\pi}
\nonumber\\&&\hspace{-14cm}
+\Delta B^{0}(m_{\pi}^2,m_{K}^2,t)(\frac{7}{6}E_{K}^2-\frac{1}{3}E_{\pi}^2-\frac{5}{6}E_{K}E_{\pi}
                                         +\frac{5}{6}m_{K}^2+\frac{1}{6}m_{\pi}^2-\frac{5}{6}t)
\nonumber\\&&\hspace{-14cm}
+\Delta B^{0}(m_{K}^2,m_{\eta}^2,t)(E_{K}^2-\frac{1}{2}E_{\pi}^2-\frac{1}{2}E_{K}E_{\pi}
                                         +\frac{7}{6}m_{K}^2-\frac{1}{6}m_{\pi}^2-\frac{1}{2}t)
\nonumber\\&&\hspace{-14cm}
-\Delta B(m_{\pi}^2,m_{K}^2,t)(\frac{5}{12}m_{K}^2+\frac{1}{12}m_{\pi}^2-\frac{5}{12}t)(E_{K}-E_{\pi})
\nonumber\\&&\hspace{-14cm}
-\Delta B(m_{K}^2,m_{\eta}^2,t)(\frac{7}{12}m_{K}^2-\frac{1}{6}m_{\pi}^2-\frac{1}{4}t)(E_{K}-E_{\pi}) )
\nonumber\\&&\hspace{-14cm}
+\frac{E_{K}-E_{\pi}}{\sqrt 2 (E_{\pi}p_{K}-E_{K}p_{\pi})F_{\pi}^2}
(  [-\frac{1}{24}\Delta A(m_{\pi},L)+\frac{7}{8} \Delta A(m_{\eta},L)+\frac{4}{3}\Delta A(m_{K},L)]p_{K}
\nonumber\\&&\hspace{-14cm}
+[\frac{29}{24}\Delta A(m_{\pi},L)-\frac{1}{8}\Delta A(m_{\eta},L)+\frac{7}{12}\Delta A(m_{K},L)]p_{\pi}
-[\frac{7}{3}\Delta B^{xx}(m_{\pi}^2,m_{K}^2,t)
\nonumber\\&&\hspace{-14cm}
+2\Delta B^{xx}(m_{K}^2,m_{\eta}^2,t)]p_{K}
-[\frac{2}{3}\Delta B^{xx}(m_{\pi}^2,m_{K}^2,t)+\Delta B^{xx}(m_{K}^2,m_{\eta}^2,t)]p_{\pi}
\nonumber\\&&\hspace{-14cm}
+\Delta B^{x}(m_{\pi}^2,m_{K}^2,t)(\frac{7}{6}p_{K}^2-\frac{1}{3}p_{\pi}^2-\frac{5}{6}p_{K}p_{\pi}
                                         +\frac{5}{6}m_{K}^2+\frac{1}{6}m_{\pi}^2-\frac{5}{6}t)
\nonumber\\&&\hspace{-14cm}
+\Delta B^{x}(m_{K}^2,m_{\eta}^2,t)(p_{K}^2-\frac{1}{2}p_{\pi}^2-\frac{1}{2}p_{K}p_{\pi}
                                         +\frac{7}{6}m_{K}^2-\frac{1}{6}m_{\pi}^2-\frac{1}{2}t)
\nonumber\\&&\hspace{-14cm}
-\Delta B(m_{\pi}^2,m_{K}^2,t)(\frac{5}{12}m_{K}^2+\frac{1}{12}m_{\pi}^2-\frac{5}{12}t)(p_{K}-p_{\pi})
\nonumber\\&&\hspace{-14cm}
-\Delta B(m_{K}^2,m_{\eta}^2,t)(\frac{7}{12}m_{K}^2-\frac{1}{6}m_{\pi}^2-\frac{1}{4}t)(p_{K}-p_{\pi}) )\,,
\end{eqnarray}
where, the momentum transfer, $t$ is
\begin{eqnarray}
t = (E_{K}-E_{\pi})^2-(p_{K}-p_{\pi})^2\,.
\end{eqnarray}
The Feynman integrals in finite volume needed to compute $\Delta f_{+}(0)$ numerically,
are obtained in the Appendix A.

\section{Numerical results}
 \label{formfactor-fv}
 \begin{figure}
 \begin{center}
 \resizebox{0.7\textwidth}{!}{%
 \includegraphics[angle=-90]{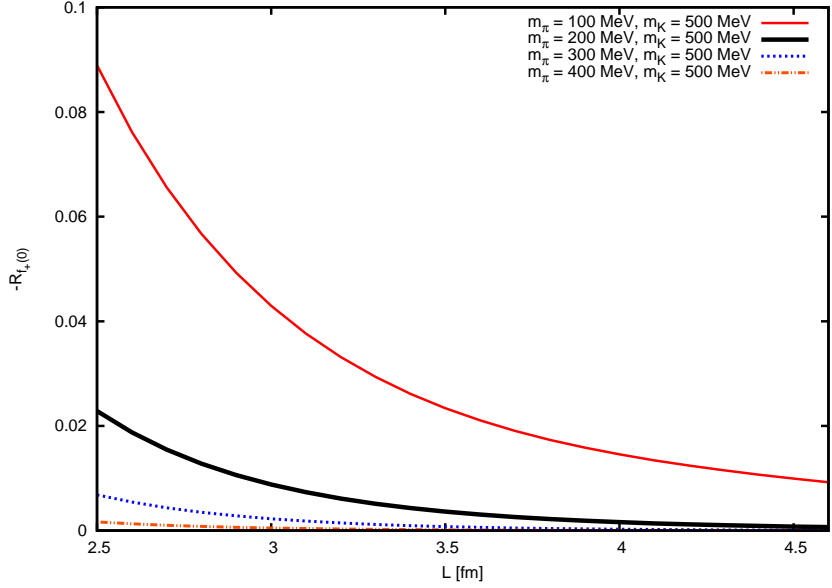}}
 \caption{The ratio $R_{f_{+}(0)}$ is plotted versus the spatial size of
 the volume, L, for three different values of pion masses
 with fixed kaon mass.}
 \label{ratio}
 \end{center}
 \end{figure}
 \begin{figure}
 \begin{center}
 \resizebox{0.7\textwidth}{!}{%
 \includegraphics[angle=-90]{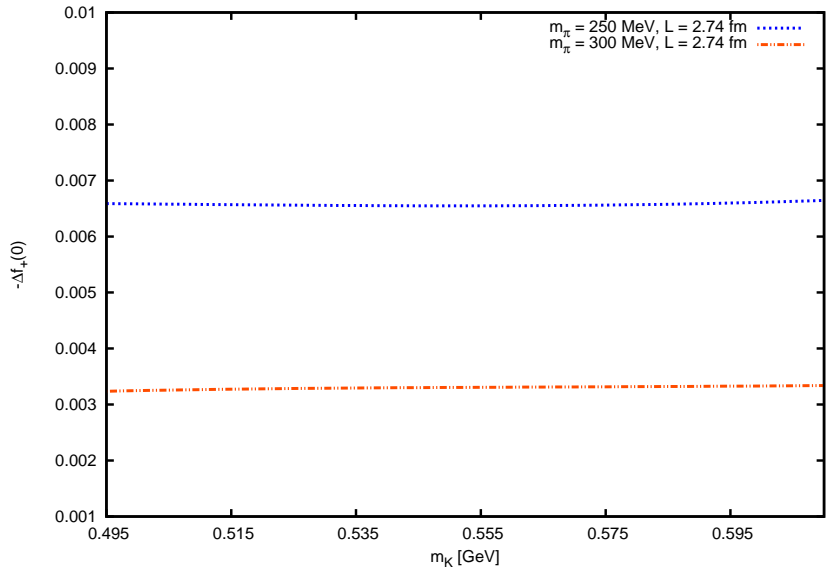}
 }
 \caption{Finite volume correction of the form factor, $\Delta f_{+}(0)$ is plotted as a function
 of the kaon mass for two different values of pion mass.}
 \label{mass}
 \end{center}
 \end{figure}

\begin{table}
\begin{center}
 \footnotesize{
 \begin{tabular}{cccccccc}
\hlinewd{.9pt}
$m_{\pi}$&$m_{K}$&    $f_{+}^{V}(0)$         &     $f_{+}^{V}(0)$      & $f^{\infty}_{+}(0)$ & $\Delta f_{+}(0)$  & $f_{+}^{V}(0)$ \\
 (GeV)   &  (GeV)& (Pole)\cite{Boyle:2007qe} &(Quadratic) \cite{Boyle:2007qe}& (ChPT $p^4$) &                  &    \\
\hlinewd{.9pt}
0.428     &  0.614  &      0.9889 (26)         &     0.9866 (33)     &    0.99315  &   -0.00980    &    0.98335       \\
0.557     &  0.666  &      0.9951 (6)         &     0.9959 (9)      &    0.99760   &   -0.00241   &     0.99518        \\
0.674     &  0.723  &      0.99925 (8)         &     0.99938 (12)    &    0.99950  &   -0.00049    &    0.99902        \\
\hlinewd{.9pt}
\end{tabular}
}
\caption{\label{fv1}
The vector form factor for different pion masses in a finite volume
with spatial size L = 1.83 fm shown in the last column is compared with the same quantity
in the third and forth columns, taken from \cite{Boyle:2007qe}, evaluated within lattice QCD
using pole and quadratic fit, respectively.}
\end{center}
\end{table}

At this point we present our numerical results. In the calculations involving $f_{+}(0)$
in infinite volume, we use for the pion decay constant its physical value
$F_{\pi} = 0.0924$ GeV.
Since we look at the vector form factor at zero momentum transfer,
the low energy constant $L^{r}_{9}$ does not show up in our expression as it
is evident from the one loop expression for $f_{+}(0)$ in Eq.~\ref{fzero}.
We show in Fig.~\ref{ratio} the numerical results for the ratio $R_{f_{+}(0)}$,
defined as
\begin{eqnarray}
R_{f_{+}(0)} = \frac{f_{+}^{V}(0)-f_{+}^{\infty}(0)}{f_{+}^{\infty}(0)} = \frac{\Delta f_{+}(0)}{f_{+}^{\infty}(0)},
\end{eqnarray}
in terms of the linear size of the volume, L.
We evaluate the vector form factor $f_{+}^{V}(0)$ in the kaon rest frame in which $p_{K} = 0$.
To obtain the pion momentum we solve the equation $t = (E_{K}-E_{\pi})^2-(p_{K}-p_{\pi})^2 = 0$ and
find $p_{\pi} = (m_{K}^2-m_{\pi}^2)/2m_{K}$.
In Fig.~\ref{ratio}, different lines stand for different choices of the pion mass
i.e. $m_{\pi} =$ 0.2 GeV, $m_{\pi} = 0.3$ GeV and $m_{\pi} = 0.4$ GeV, while
a fixed value of the kaon mass, $m_{K} = 0.5$ GeV, is chosen.
We apply the GMOR relation at leading order, $m_{\eta}^2 = (4 m_{K}^2-m_{\pi}^2)/3$, to
obtain the eta mass needed in our calculation.
The results in Fig.~\ref{ratio} indicate two standard characteristic features since the ratio
tends to zero asymptotically and on top of that the
ratio grows with decreasing the pion mass.
We found out in \cite{Kgh2011} that the finite size effects of the scalar form factor
at the maximum momentum transfer becomes larger for larger pion mass.
The reason for this unusual feature was because of polynomial terms in
front of the $B^{0}$ function which its growth with pion mass is fast.
For our case in the present article, the vector form factor at $q^2=0$,
no such terms show up and therefore we see the natural expectation
where, the finite volume effects are smaller for larger pion mass.

It is noted in the literature, see for example \cite{Boyle:2010bh},
that the strange quark mass acts as a regulator of the finite volume effects.
In order to examine this fact here, we use the GMOR relation and plot $\Delta f_{+}(0)$
as a function of kaon mass for two values of pion mass namely,
$m_{\pi} = 0.250$ GeV and $m_{\pi} = 0.300$ GeV. Our result depicted in Fig.~\ref{mass}
confirms the current knowledge that the impact of the heavy Pseudo-Goldstone boson, namely,
eta and kaon loops on the finite volume effects are basically unimportant.

Moreover, we have calculated the vector form factor at zero momentum transfer
for two ensembles corresponding to volumes with $L = 1.83$ fm and $L = 2.74$ fm
as quoted in \cite{Boyle:2007qe}.
For the smaller volume with $L = 1.83$ fm, in Table.~\ref{fv1} finite volume corrections of
the vector form factor are presented at zero momentum transfer.
Given the vector form factor at infinite volume at order $p^4$, we can obtain the vector
form factor at finite volume. We compare our results with the lattice data in
\cite{Boyle:2007qe} and see that they are somewhat close to the lattice data.
We also compute the form factor at a volume with $L = 2.74$ fm and compare with lattice data
in \cite{Boyle:2007qe} provided by Table.~\ref{fv2}.
\begin{table}
\begin{center}
 \footnotesize{
 \begin{tabular}{cccccccc}
\hlinewd{.9pt}
$m_{\pi}$& $m_{K}$ & $            f_{+}^{V}(0)$ & $f_{+}^{V}(0)$                 & $f^{\infty}_{+}(0)$ &  $\Delta f_{+}(0)$  & $f_{+}^{V}(0)$ \\
 (GeV)   &   (GeV) & (Pole) \cite{Boyle:2007qe} & (Quadratic) \cite{Boyle:2007qe} &  (ChPT $p^4$)   &                     & \\
\hlinewd{.9pt}
0.329    &  0.575   &      0.9774 (35)    &     0.9749 (59)    &    0.98825   &  -0.00258     &    0.98567       \\
0.416    &  0.604   &      0.9841 (29)    &     0.9806 (39)    &    0.99301   &  -0.00087     &    0.99214        \\
0.556    &  0.663   &      0.9960 (7)     &     0.9962 (9)      &    0.99769  &  -0.00013     &    0.99756        \\
0.671    &  0.719   &      0.9991 (2)     &     0.9990 (2)      &    0.99952  &  -0.00002     &    0.99950       \\
\hlinewd{.9pt}
\end{tabular}
}
\caption{\label{fv2}
The vector form factor for different pion masses in a finite volume
with spatial size L = 2.74 fm shown in the last column is compared with the same quantity
in the third and forth columns, taken from \cite{Boyle:2007qe}, evaluated
within lattice QCD using pole and quadratic fit, respectively.}
\end{center}
\end{table}

In order to compare our finite volume effects with ones from lattice
data in \cite{Boyle:2007qe}, we define the
quantity $\Delta = f^{V_{1}}_{+}(0)-f^{V_{2}}_{+}(0)$ and
present in Table.~\ref{fve} its numerical values for two
volumes with $L_{1} = 2.74$ fm and $L_{2} = 1.83$ fm.
The quantity $\Delta$ does not depend on the form factors
in infinite volume.
The three sets of pion and kaon masses shown in Table.~\ref{fve}
are almost the same in two different volumes where
we chose their values in volume with $L = 1.83$ fm.
\begin{table}
\begin{center}
 \footnotesize{
 \begin{tabular}{ccccc}
\hlinewd{.9pt}
$m_{\pi}$& $m_{K}$ &   $\Delta$                 &  $\Delta$                        & $\Delta$         \\
 (GeV)   &   (GeV) & (Pole) \cite{Boyle:2007qe} & (Quadratic) \cite{Boyle:2007qe} & ${\cal O}(p^4$)    \\
\hlinewd{.9pt}
0.428    &  0.614   &   -0.0048   &   -0.006     &    0.0089       \\
0.557    &  0.666   &    0.0009   &    0.0003    &    0.00228       \\
0.674    &  0.723   &   -0.00015  &   -0.00038   &    0.00047       \\
\hlinewd{.9pt}
\end{tabular}
}
\caption{\label{fve}
The quantity $\Delta$ is evaluated by using lattice data
in \cite{Boyle:2007qe} and compared with ones from
ChPT results for three sets of pion and kaon masses
in two volumes with $L_{1} = 2.74$ fm and $L_{2} = 1.83$ fm.}
\end{center}
\end{table}
Given the fact that the size of the finite volume
correction is larger for smaller volume, we expect to have $\Delta > 0$.
We agree in the sign of $\Delta$ with lattice data
only for a set with $m_{\pi} = 0.557$ GeV even though,
our ChPT results predict a bigger value for $\Delta$.
For the other two sets, the size of $\Delta$ from
ChPT results are comparable with ones from lattice
QCD with quadratic fit, but we disagree in the sign
of $\Delta$.

Recently, RBC-UKQCD released their data within domain wall lattice QCD with three
dynamical quark flavors in \cite{Boyle:2013}. We compare our ChPT results with
their data in two volumes in Table.~\ref{fv3}.

\begin{table}
\begin{center}
 \footnotesize{
 \begin{tabular}{cccccccc}
\hlinewd{.9pt}
$m_{\pi}$& $m_{K}$ &  L  & $  f_{+}^{V}(0)$ &  $f^{\infty}_{+}(0)$      &  $\Delta f_{+}(0)$  & $f_{+}^{V}(0)$ \\
 (GeV)   &   (GeV) &(fm) &\cite{Boyle:2013} &   (ChPT $p^4$)   &                     & \\
\hlinewd{.9pt}
0.171    &  0.493   &  4.56  &  0.9710 (45) &    0.98086   &  -0.00143    &    0.97943       \\
0.248    &  0.510   &  4.59 &  0.9771 (21)  &    0.98688   &  -0.00021    &    0.98667      \\
0.334    &  0.580   &  2.75  &  0.9760 (43) &    0.98824  &  -0.00233    &    0.98591      \\
0.563    &  0.671   &  2.73  &  0.9956 (4)  &    0.99764  &  -0.00012     &    0.99752     \\
0.678    &  0.726   &  2.74  &  0.9992 (1)  &    0.99952  &  -0.00002     &    0.99950        \\
\hlinewd{.9pt}
\end{tabular}
}
\caption{\label{fv3}
The vector form factor given in the last column for different pion masses in finite volume
is compared with the same quantity
in the forth column, taken from \cite{Boyle:2013}, computed
within domain wall lattice QCD with $N_{f}=2+1$.}
\end{center}
\end{table}
In addition, there are lattice data for the form factor from ETM collaboration in \cite{Lubicz2009},
obtained from simulations with two flavors of dynamical twisted-mass fermions.
We compute the form factor for two ensembles as noted in \cite{Lubicz2009} and
compare with the corresponding lattice data in Table.~\ref{fv4}.
\begin{table}
\begin{center}
 \footnotesize{
 \begin{tabular}{ccccccccc}
\hlinewd{.9pt}
$m_{\pi}$& $m_{K}$ &  L  & $  f_{+}^{V}(0)$         &  $  f_{+}^{V}(0)$     &  $f^{\infty}_{+}(0)$  &  $\Delta f_{+}(0)$  & $f_{+}^{V}(0)$ \\
 (GeV)   &   (GeV) &(fm) &(Pole) \cite{Lubicz2009}  & (Quadratic) \cite{Lubicz2009}   &   (ChPT $p^4$)        &                     & \\
\hlinewd{.9pt}
0.300    &  0.530   &  2.133  &  0.98633 (362) &  0.98597 (337)   &   0.98974    &  -0.01599    &    0.97377       \\
0.300    &  0.530   &  2.83   &  0.98052 (440) &  0.97950 (390)   &   0.98974    &  -0.00316    &    0.98658      \\
\hlinewd{.9pt}
\end{tabular}
}
\caption{\label{fv4}
The vector form factor given in the last column for different pion masses in finite volume
is compared with the same quantity
in the forth and fifth column, taken from \cite{Lubicz2009}, computed
within dynamical twisted-mass fermions lattice QCD with $N_{f}=2$.}
\end{center}
\end{table}
We know that finite volume correction becomes smaller for larger volumes while keeping pion and kaon
masses fixed. Therefore, as it is evident from our results in Table.~\ref{fv4}, the form factor
in finite volume is larger for the smaller volume. This behavior cannot be seen in the lattice data
shown in Table.~\ref{fv4}.

\section{Conclusion}
\label{summary}
In this work we have found an analytical expression for the
finite volume correction of the kaon semi-leptonic
vector form factor at momentum transfer $t$, which is called $\Delta f^{V}_{+}(t)$.
We have presented the numerical estimates of $\Delta f^{V}_{+}(0)$
and studied its dependence on the pion mass and kaon mass.
We emphasize on the fact that varying the kaon mass and correspondingly
the eta mass have a very tiny effects on the finite
volume corrections in comparison with the effects which emanate from
the variation of the pion mass.

Moreover, we compute $f^{V}_{+}(0)$ numerically and compare with lattice data.
Our vector form factor in finite volume
consists of two parts. One is the vector form factor in infinite volume which
we use its values at one-loop order from ChPT and the second part is the finite
box effects which are evaluated in this article at one-loop order.
Even at this order of computation, we see that our results are comparable to some extend
with the lattice data in \cite{Boyle:2013,Boyle:2007qe}. However, there are possible sources
of improvements to our results. First of all, one expects some modifications in our results
due to the two-loop corrections of the first part, i.e., the vector form factor
in infinite volume, even thought these corrections are not fully predicted by ChPT.
The other possible improvement we can make in our results arise from the two-loop
finite volume corrections. In fact, the order $p^6$ finite volume effects do not depend
on the order $p^6$ coupling constants, but they are not free from the order $p^4$
coupling constants. One may therefore extend the present finite volume
corrections to two-loop order precision.

We have also computed the quantity $\Delta$ which compares the size
of the finite volume effects in two different volumes. Our ChPT results
and those from lattice data in \cite{Boyle:2007qe} are collected in
Table.~\ref{fve}. We realize that in two sets of masses our predicted
values for $\Delta$ are comparable in size with the corresponding values from
lattice data but disagree in the sign of $\Delta$. We also notice form
lattice QCD predictions that the sign of $\Delta$ are not the same for
all three sets of masses which is in contrast to our expectation.
It is therefore difficult to draw a concrete
conclusion from comparisons done in Table.~\ref{fve}.

\renewcommand{\theequation}{A-\arabic{equation}}
\setcounter{equation}{0}  

\section{Appendix A}
\label{FVcalculation}
In this part we calculate the Feynman integrals in finite volume
for a generic momentum transfer $q^2$. There are two types of integrals
which appear in our expressions for form factors: scalar integrals
and tensor integrals. In fact we wish to evaluate finite volume correction
for a given integral for which we define $\Delta I = I_{V}- I_{\infty}$, where
subscripts $\infty$ and $V$ indicate integration in infinite and finite
volume, respectively.

\subsection{One loop scalar integrals}

The simplest integral we encounter in this work is related
to the tadpole Feynman diagram
\begin{eqnarray}
A(M^2) =  \frac{1}{i} \int \frac{d^dp}{(2\pi)^d}
\frac{1}{p^2-M^2}.
\end{eqnarray}
In finite volume, momentum gets quantized and therefore integration
over momentum is replaced by summation
\begin{eqnarray}
\label{afunc}
A_{V} (M^2)  = -\frac{i}{L^3} \sum_{\vec{p}} \int \frac{dp_0}{2\pi}
\frac{1}{p^2-M^2)}
\nonumber\\&&\hspace{-5.3cm}
= A_{\infty}-\int\frac{dp_0}{2\pi}\sum_{\vec{n}\neq 0}\int \frac{d^{3}\vec{p}}{(2\pi)^3}
           \frac{i e^{iL\vec{p}.\vec{n}}}{p^2-M^2},
\end{eqnarray}
where $A_{\infty}$ is the value of the integral in infinite volume.
In obtaining the second line, we have employed the Poisson summation formula.
By taking a contour integration over $p_{0}$ and then performing the three
dimensional integral we achieve the known result \cite{Kgh2006}
\begin{eqnarray}
\Delta A = - \frac{M}{4 \pi^2 L} \sum_{\vec{n} = \vec{1}}
\frac{1}{\vert \vec{n} \vert} m(n) K_{1}(M L \vert \vec{n}\vert),
\end{eqnarray}
where, $K_{1}$ is the modified Bessel function of order one
and the multiplicity factor $m(n)$ stands for the number of
possibilities that the relation $n = n_{1}^2+n_{2}^2+n_{3}^2$
is satisfied for a given value of $n$ with positive and negative
integer numbers of $n_{1}$, $n_{2}$ and $n_{3}$. $m(n)$ factors are
listed in Table.~\ref{multy}.
\begin{table}
\begin{center}
\end{center}
 \footnotesize{
    \begin{tabular}{ccccccccccccccccccccc}
\hlinewd{.9pt}
n & 1 & 2 & 3 & 4 & 5 & 6 & 7 & 8 & 9 & 10 & 11 & 12 & 13 & 14 & 15 & 16 & 17 & 18 & 19 & 20  \\
\hlinewd{.9pt}
m(n) & 6 & 12 & 8 & 6 & 24 & 24 & 0 & 12 & 30 & 24 & 24 & 8 & 24 & 48 & 0 & 6 & 48 & 36 & 24 & 24 \\
\hlinewd{.9pt}
\end{tabular}
}
\caption{\label{multy}
The multiplicity factors $m(n)$ are provided for $1 \le n \le 20$.}
\end{table}

The next integral we should evaluate in finite volume as a new one
is related to the rescattering effects at momentum transfer $q^2$
\begin{eqnarray}
B(m^2,M^2,q^2) =  \frac{1}{i} \int \frac{d^dp}{(2\pi)^d}
\frac{1}{(p^2-m^2)((q+p)^2-M^2)},
\end{eqnarray}
where we assume hereafter that $M > m$.
With the application of the Poisson summation formula
\begin{table}
\begin{center}
 \footnotesize{
    \begin{tabular}{cccc}
\hlinewd{.9pt}
n        &   $C_{n}(\alpha x)$   &  n    &  $C_{n}(\alpha x)$   \\
\hlinewd{.9pt}
1       &   $2\cos(\alpha x)+4$    &  11   &  $8\cos(3\alpha x)+16\cos(\alpha x)$                    \\
2       &   $8\cos(\alpha x)+4$ &  12   &  $8\cos(2\alpha x)$  \\
3       &   $8\cos(\alpha x)$ & 13  &  $8\cos(3\alpha x)+8\cos(2\alpha x)+8$    \\
4   &     $2\cos(2\alpha x)+4$  & 14  &  $16\cos(3\alpha x)+16\cos(2\alpha x)+16\cos(\alpha x)$  \\
5   &     $8\cos(2\alpha x)+8\cos(\alpha x)+8$  & 15 & 0      \\
6   &     $8\cos(2\alpha x)+16\cos(\alpha x)$  & 16 & $2\cos(4\alpha x)+4$   \\
7   &  0       & 17           &  $8\cos(4\alpha x)+8\cos(3\alpha x)+$     \\
    &          &              &  $16\cos(2\alpha x)+8\cos(\alpha x)+8$    \\
8   &  $8\cos(2\alpha x)+4$  & 18  &  $6\cos(4\alpha x)+10\cos(3\alpha x)$ \\
    &                       &     &  $+16\cos(\alpha x)+4$    \\
9   &  $2\cos(3\alpha x)+16\cos(2\alpha x)+8\cos(\alpha x)+4$  & 19 &  $18\cos(3\alpha x)+6\cos(\alpha x)$     \\
10   &  $8\cos(3\alpha x)+8\cos(\alpha x)+8$  & 20 & $8\cos(4\alpha x)+8\cos(2\alpha x)+8$     \\
\hlinewd{.9pt}
\end{tabular}
}
\caption{\label{functions1}
Functions $C_{n}(\alpha x)$ are provided for $1 \le n \le 20$ when external momentum is chosen as $\vec{q} = q_{x}~(1,0,0)$.}
\end{center}
\end{table}
and making use of the Feynman parameter formula followed by
redefining the variable $p_{0}$ we will arrive at
\begin{eqnarray}
\Delta B (m^2,M^2,q^2) =
-i \int_{0}^{1}dx \int \frac{dp_0}{2\pi} \sum_{\vec{n}\neq 0}
\nonumber\\&&\hspace{-4.8cm}
\int \frac{d^{3}\vec{p}}{(2\pi)^3}
\frac{e^{iL\vec{p}.\vec{n}}}{\Large[ p_{0}^2-(\vec{p}+(1-x)\vec{q}~)^2+x(1-x)q^2-xm^2-(1-x)M^2 \Large]^2}.
\nonumber\\&&\hspace{-4.8cm}
\end{eqnarray}
At the next step, we begin by taking the contour integral over $p_{0}$
and then make a redefinition of the variable $\vec{p}$ to obtain
\begin{eqnarray}
\Delta B (m^2,M^2,q^2) =
\frac{1}{4}\sum_{\vec{n}\neq 0} \int_{0}^{1} dx~e^{-iLx\vec{q}.\vec{n}}
\int \frac{d^{3}\vec{p}}{(2\pi)^3}
\frac{e^{iL\vec{p}.\vec{n}}}{[\vec{p}^2-x(1-x)q^2+xm^2+(1-x)M^2]^{3/2}}.
\nonumber\\&&\hspace{-4.8cm}
\end{eqnarray}
The exponential factor $e^{-iLx\vec{q}.\vec{n}}$
explicitly breaks the rotational symmetry in the expression above.
We carry out the integral over the vector
momentum in two steps. We take first an integral
over the angular part of the three dimensional momentum and then
we make use of the convolution technique to perform the
final integral. We find the following result
\begin{eqnarray}
\label{Bfunc}
\Delta B(m^2,M^2,q^2) =  \frac{1}{8\pi^2} \sum_{\vec{n} \neq 0}
\int_{0}^{1} dx~ C_{n}(\alpha x)~K_{0}(wQ)\,,
\end{eqnarray}
where $w = L |\vec{n}|$ and $Q = \sqrt{xm^2+(1-x)M^2-x(1-x)q^2}$.
$K_{0}$ is the modified Bessel function of rank zero.
This is a generalization of the case with $M = m$ obtained
in \cite{Kgh2012}.
Functions $C_{n}(\alpha x)$ introduced
in the expression above involve the exponential factor $e^{-iLx\vec{q}.\vec{n}}$, where
we have summed over all possible ways that for a given $n$
the relation $n = n_{1}^2+n_{2}^2+n_{3}^2$ is satisfied.
In this article we take the momentum transfer, $\vec{q}$, along the $x$-axis,
i.e, $\vec q = (q_{x},0,0)$. We provide functions $C_{n}(\alpha x)$
in Table~\ref{functions1}, where $\alpha = L q_{x}$.
It is easy to see that for $\alpha = 0$, functions $C_{n}(\alpha x)$
are identical to multiplicity factors $m(n)$.

\subsection{One loop tensor integrals}
In this section we consider the tensor integrals by calculating
their temporal and spatial components.
We begin with the temporal component of the tensor integrals.
The next integral we then need to consider in finite volume is
\begin{eqnarray}
B^{0}(m^2,M^2,q^2) =  \frac{1}{i} \int \frac{d^dp}{(2\pi)^d}
\frac{p_{0}}{(p^2-m^2)((q+p)^2-M^2)}.
\end{eqnarray}
$B^{0}$ is the temporal component of the tensor integral and $B^{\mu}$
is defined in Appendix B. By repeating the procedures
stated above we can readily prove that
\begin{eqnarray}
\label{b0func}
\Delta B^{0}(m^2,M^2,q^2) = -\frac{q_{0}}{8\pi^2} \sum_{\vec{n} \neq 0}
\int_{0}^{1} dx~x~C_{n}(\alpha x)~K_{0}(wQ)\,,
\end{eqnarray}
where, $q_{0} = E_{p^\prime}-E_{p}$.
\begin{table}
\begin{center}
 \footnotesize{
    \begin{tabular}{cccc}
\hlinewd{.9pt}
n        &   $D_{n}(\alpha x)$   &  n    &  $D_{n}(\alpha x)$   \\
\hlinewd{.9pt}
1       &   $2\sin(\alpha x)$    &  11   &  $8\sin(3\alpha x)+16\sin(\alpha x)$                    \\
2       &   $8\sin(\alpha x)$ &  12   &  $8\sin(2\alpha x)$  \\
3       &   $8\sin(\alpha x)$ & 13  &  $8\sin(3\alpha x)+8\sin(2\alpha x)$    \\
4   &     $2\sin(2\alpha x)$  & 14  &  $16\sin(3\alpha x)+16\sin(2\alpha x)+16\sin(\alpha x)$  \\
5   &     $8\sin(2\alpha x)+8\sin(\alpha x)$  & 15 & 0      \\
6   &     $8\sin(2\alpha x)+16\sin(\alpha x)$  & 16 & $2\sin(4\alpha x)$   \\
7   &  0       & 17           &  $8\sin(4\alpha x)+8\sin(3\alpha x)$     \\
    &          &              &  $+16\sin(2\alpha x)+8\sin(\alpha x)$    \\
8   &  $8\sin(2\alpha x)$  & 18  &  $6\sin(4\alpha x)+10\sin(3\alpha x)+16\sin(\alpha x)$    \\
9   &  $2\sin(3\alpha x)+16\sin(2\alpha x)+6\sin(\alpha x)$  & 19 &  $18\sin(3\alpha x)+6\sin(\alpha x)$     \\
10   &  $8\sin(3\alpha x)+8\sin(\alpha x)$  & 20 & $8\sin(4\alpha x)+8\sin(2\alpha x)$     \\
\hlinewd{.9pt}
\end{tabular}
}
\caption{\label{functions2}
Functions $D_{n}(\alpha x)$ are provided for $1 \le n \le 20$ when external momentum is chosen as $\vec{q} = q_{x}~(1,0,0)$.}
\end{center}
\end{table}
Now we look at the temporal component of the tensor integral $B^{\mu \nu}$
\begin{eqnarray}
\label{b00integral}
B^{00}(m^2,M^2,q^2) =  \frac{1}{i} \int \frac{d^dp}{(2\pi)^d}
\frac{p_0^2}{(p^2-m^2)((q+p)^2-M^2)}.
\end{eqnarray}
We redo the procedures sketched above and finally arrive at
\begin{eqnarray}
\label{b00func}
\Delta B^{00}(m^2,M^2,q^2)  =  -\frac{1}{8\pi^2 L} \sum_{\vec{n} \neq 0}
\frac{1}{|\vec n|} \int_{0}^{1} dx
~C_{n}(\alpha x)~Q~K_{1}(wQ)
\nonumber\\&&\hspace{-8.5cm}
+\frac{q_{0}^2}{8\pi^2 L} \sum_{\vec{n} \neq 0}
\frac{1}{|\vec n|} \int_{0}^{1} dx~x^2~C_{n}(\alpha x)~K_{0}(wQ).
\end{eqnarray}
Moreover, the tensor integral $B^{\mu}$ has a spatial component
\begin{eqnarray}
B^{x}(m^2,M^2,q^2) =  \frac{1}{i} \int \frac{d^dp}{(2\pi)^d}
\frac{p_{x}}{(p^2-m^2)((q+p)^2-M^2)}.
\end{eqnarray}
In order to find the integral $B^{x}$ in finite volume we follow the same path
as we did to evaluate $B^{0}$. Our final result reads
\begin{eqnarray}
\label{bxfunc}
\Delta B^{x}(m^2,M^2,q^2) = \frac{1}{8\pi^2} \sum_{\vec{n} \neq 0}
\int_{0}^{1} dx~ \frac{D_{n}(\alpha x)}{|\vec n|}~Q~K_{1}(wQ)
\nonumber\\&&\hspace{-7cm}
-\frac{q_{x}}{8\pi^2} \sum_{\vec{n} \neq 0}\int_{0}^{1} x~dx\times
C_{n}(\alpha x)~K_{0}(wQ).
\end{eqnarray}
Functions $D_{n}(\alpha x)$ appear when for a given $n$,
we compute the summation $\sum i~n_{x} e^{-iLx\vec{q}.\vec{n}}$ over
all possible values of $(n_{1},n_{2},n_{3})$ that fulfill the relation
$n = n_{1}^2+n_{2}^2+n_{3}^2$. Functions $D_{n}(\alpha)$ are listed
in Table~\ref{functions2}.\\
\begin{table}
\begin{center}
 \footnotesize{
    \begin{tabular}{cccc}
\hlinewd{.9pt}
n        &   $F_{n}(\alpha x)$   &  n    &  $F_{n}(\alpha x)$   \\
\hlinewd{.9pt}
1       &   $2\cos(\alpha x)$    &  11   &  $72\cos(3\alpha x)+16\cos(\alpha x)$             \\
2       &   $8\cos(\alpha x)$ &  12   &  $32\cos(2\alpha x)$  \\
3       &   $8\cos(\alpha x)$ & 13  &  $72\cos(3\alpha x)+32\cos(2\alpha x)$    \\
4   &     $8\cos(2\alpha x)$  & 14  &  $144\cos(3\alpha x)+64\cos(2\alpha x)+16\cos(\alpha x)$  \\
5   &     $32\cos(2\alpha x)+8\cos(\alpha x)$  & 15 & 0      \\
6   &     $32\cos(2\alpha x)+16\cos(\alpha x)$  & 16 & $32\cos(4\alpha x)$   \\
7   &  0       & 17           &  $128\cos(4\alpha x)+72\cos(3\alpha x) $ \\
    &          &              &  $+64\cos(2\alpha x)+8\cos(\alpha x)$    \\
8   &  $32\cos(2\alpha x)$  & 18  &  $96\cos(4\alpha x)+90\cos(3\alpha x)+16\cos(\alpha x)$    \\
9   &  $18\cos(3\alpha x)+64\cos(2\alpha x)+6\cos(\alpha x)$  & 19 &  $162\cos(3\alpha x)+6\cos(\alpha x)$     \\
10   &  $72\cos(3\alpha x)+8\cos(\alpha x)$  & 20 & $128\cos(4\alpha x)+32\cos(2\alpha x)$     \\
\hlinewd{.9pt}
\end{tabular}
}
\caption{\label{functions3}
Functions $F_{n}(\alpha x)$ are provided for $1 \le n \le 20$ when external momentum is chosen as $\vec{q} = q_{x}~(1,0,0)$.}
\end{center}
\end{table}
The last integral in finite volume is the spatial
component $B^{xx}$ of the tensor integral $B^{\mu \nu}$
defined in the Appendix B,
\begin{eqnarray}
B^{xx}(m^2,M^2,q^2) =  \frac{1}{i} \int \frac{d^dp}{(2\pi)^d}
\frac{p_{x}^{2}}{(p^2-m^2)((q+p)^2-M^2)}.
\end{eqnarray}
For the finite volume correction of the integral above we obtain
\begin{eqnarray}
\label{bxxfunc}
\Delta B^{xx}(m^2,M^2,q^2) = \frac{1}{8\pi^2} \sum_{\vec{n} \neq 0}
\int_{0}^{1} dx~  \Large[ -\frac{F_{n}(\alpha x)}{L|\vec n|^3}~Q~K_{1}(wQ)
\nonumber\\&&\hspace{-9cm}
-\frac{F_{n}(\alpha x)}{2|\vec n|^2} Q^2[K_{0}(wQ)
+K_{2}(wQ)]+ \frac{C_{n}(\alpha x)}{L|\vec n|}~K_{1}(wQ) \Large]
\nonumber\\&&\hspace{-9cm}
+\frac{q_{x}^2 }{8\pi^2} \sum_{\vec{n} \neq 0}
\int_{0}^{1} x^2~dx~C_{n}(\alpha x)~K_{0}(wQ)
\nonumber\\&&\hspace{-9cm}
-\frac{q_{x}}{4\pi^2} \sum_{\vec{n} \neq 0}
\int_{0}^{1} x~dx~\frac{D_{n}(\alpha x)}{|\vec n|}~Q~K_{1}(wQ),
\end{eqnarray}
Functions $F_{n}(\alpha x)$ listed in Table.~\ref{functions3},
are obtained by computing
$\sum n_{x}^2 e^{-iLx\vec{q}.\vec{n}}$ for a given $n$
over all possible ways that the relation $n = n_{1}^2+n_{2}^2+n_{3}^2$ holds.

\renewcommand{\theequation}{B-\arabic{equation}}
\setcounter{equation}{0}  

\section{Appendix B}
We introduce the necessary one loop scalar Feynman integrals for the discussed
decay
\begin{eqnarray}
A(m^2) =  \frac{1}{i} \int \frac{d^dp}{(2\pi)^d} \frac{1}{p^2-m^2},
\end{eqnarray}
\begin{eqnarray}
B(m^2,M^2,q^2)&&\hspace{-0.1cm} =  \frac{1}{i} \int \frac{d^dp}{(2\pi)^d}
\frac{1}{(p^2-m^2)((p+q)^2-M^2)},
\end{eqnarray}
and for the tensor Feynman integrals
\begin{eqnarray}
B_{\mu}(m^2,M^2,q^2)&&\hspace{-0.1cm} =  \frac{1}{i} \int \frac{d^dp}{(2\pi)^d}
\frac{p_{\mu}}{(p^2-m^2)((p+q)^2-M^2)},
\end{eqnarray}
\begin{eqnarray}
B_{\mu \nu}(m^2,M^2,q^2)&&\hspace{-0.1cm} =  \frac{1}{i} \int \frac{d^dp}{(2\pi)^d}
\frac{p_{\mu} p_{\nu}}{(p^2-m^2)((p+q)^2-M^2)}.
\end{eqnarray}
One can write the tensor integrals
in terms of scalar functions by applying Lorentz symmetry
\begin{eqnarray}
B_{\mu}(m^2,M^2,q^2) =  q_{\mu} B_{1}(m^2,M^2,q^2) ,
\end{eqnarray}
\begin{eqnarray}
B_{\mu \nu}(m^2,M^2,q^2)&&\hspace{-0.1cm} =  q_{\mu} q_{\nu} B_{12}(m^2,M^2,q^2)
+g_{\mu \nu} B_{22}(m^2,M^2,q^2).
\end{eqnarray}

\section*{Acknowledgement}
We would like to thank Hans Bijnens for useful discussions.
K.G. acknowledges Arak University for financial support under the contract No.92/52.

\end{document}